\documentclass{aa}
\input{epsf}

\begin{document}

 \def \etal      {et al.\ }
\def \kev       {{\rm\ keV}}
\title{Measuring Cluster Temperature Profiles with XMM/EPIC\thanks{Based on observations obtained with XMM-Newton, an 
      ESA science mission with instruments and contributions directly
funded by
    ESA Member States and the USA (NASA)}
\fnmsep
\thanks{EPIC was developed by the EPIC Consortium led by the Principal
Investigator, Dr. M. J. L. Turner. The consortium comprises the
following Institutes: University of Leicester, University of
Birmingham, (UK); CEA/Saclay, IAS Orsay, CESR Toulouse, (France);
IAAP Tuebingen, MPE Garching,(Germany); IFC Milan, ITESRE Bologna,
IAUP Palermo, Italy. EPIC is funded by: PPARC, CEA, CNES, DLR and ASI
}}
\author{M. Arnaud\inst{1}, D.M. Neumann\inst{1}, N. Aghanim\inst{2},
   R. Gastaud\inst{3},
S.Majerowicz\inst{1}, John P.~Hughes\inst{1,4}}

\offprints{M. Arnaud, marnaud@discovery.saclay.cea.fr}

\institute{CEA/DSM/DAPNIA Saclay, Service d'Astrophysique, L'Orme
    des Merisiers B\^at 709., 91191 Gif-sur-Yvette, France \and
    IAS-CNRS, Universit\'{e} Paris Sud, B\^atiment 121, 91405 Orsay
    Cedex, France \and CEA/DSM/DAPNIA Saclay, Service d'Electronique
    et d'Informatique, 91191 Gif-sur-Yvette, France\and Department of
    Physics and Astronomy, Rutgers University, 136 Frelinghuysen Road,
    Piscataway, NJ 08854-8019 USA}

  \date{Received 2 October 2000 / Accepted 4 November 2000}

\titlerunning{Measuring Cluster Temperature Profiles with XMM/EPIC}
\authorrunning{Arnaud \etal}

\abstract{Using the PV observation of A1795, we illustrate the
capability of XMM-EPIC to measure cluster temperature profiles, a key
ingredient for the determination of cluster mass profiles through the
equation of hydrostatic equilibrium.  We develop a methodology for
spatially resolved spectroscopy of extended sources, adapted to XMM
background and vignetting characteristics.  The effect of the particle
induced background is discussed.  A simple unbiased method is proposed
to correct for vignetting effects, in which every photon is weighted
according to its energy and location on the detector.  We were able to
derive the temperature profile of A1795 up to 0.4 times the virial
radius.  A significant and spatially resolved drop in temperature
towards the center ($r<200~{\rm kpc}$) is observed, which corresponds
to the cooling flow region of the cluster.  Beyond that region, the
temperature is constant with no indication of a fall-off at large
radii out to $1.2$ Mpc.  \keywords{ Galaxies: clusters: individual:
A1795 -- Galaxies: intergalactic medium -- Cosmology: observations --
Cosmology: dark matter -- X-rays: general }}

\maketitle

\section{Introduction}
The determination of the temperature profiles of the hot gas in galaxy
clusters is essential to derive the gas entropy distribution and to
measure the total mass content, through the equation of hydrostatic
equilibrium.  These are key quantities for cosmological studies based
on the properties of galaxy clusters.  The baryonic mass fraction in
clusters is a strong constraint on the density parameter of the
Universe $\Omega$ (Briel, Henry \& B\"{o}hringer \cite{briel1}; White \etal
\cite{white1}).  If the mass--temperature relation is well calibrated, the mass
distribution function and its evolution can be derived from the
observed temperature function.  This is an independent and powerful
probe of $\Omega$, as well as, the index and normalization of the
power spectrum of primeval fluctuations (e.g. Oukbir \& Blanchard
\cite{oukbir}).  Furthermore, the physics of structure formation and evolution
can be assessed.  For example, the total mass profile gives
information on the physics of gravitational collapse.  Current
numerical simulations under a cold dark matter scenario predict a
universal dark matter profile (Navarro, Frenk \& White \cite{NFW}).  The
gas entropy distribution, the relative distribution of the gas and
dark matter and departures from self-similarity constrain the
thermo-dynamical history of the hot gas and the importance of
additional non-gravitational effects, such as galaxy feedback (e.g.
Ponman, Cannon \& Navarro \cite{ponman}).

While gas density profiles have been determined with good accuracy by
ROSAT, the shape of the temperature profiles in clusters, as measured
by ASCA and SAX, is still a matter of debate (Markevitch \etal
\cite{markevitch}; Irwin \etal \cite{irwin1}; White \cite{white2};
Irwin \& Bregman \cite{irwin2}; Molendi \etal \cite{molendi}; Ettori
\etal \cite{ettori}).  It is at present unclear if cluster atmospheres
are essentially isothermal or if the temperature decreases with radius
(as expected from numerical simulations of cluster formation).  As a
result, total mass estimates are typically uncertain by a factor of
two (Neumann \& Arnaud \cite{neumann1}; Horner, Mushotzky \& Scharf
\cite{horner}) and the mass profiles are too uncertain to provide any
real constraint on theoretical models.  

\begin{figure}[t]
\epsfxsize=8.cm \epsfverbosetrue \epsfbox{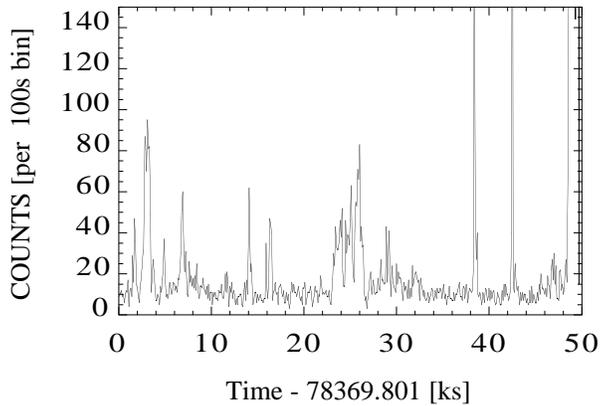} \caption[
]{{\footnotesize The total EPIC/MOS1 count rate in the $[10-12]$ keV
energy band as a function of time.  }} \label{fig:frates} \end{figure}

Significant progress was expected with XMM-Newton/EPIC (Jansen \etal
\cite{jansen}, Turner \etal \cite{turner}), which has a much better
sensitivity than ASCA and SAX and does not suffer from the large
energy dependent PSF of ASCA, a major source of systematic
uncertainty.  In this paper, we illustrate the capability of
XMM-Newton to measure cluster temperature profiles, using the PV
observation of A1795.  A1795 is a bright cluster ($S_{\rm X}[2-10]\kev
> 5\times 10^{-11}~ {\rm ergs/s/cm^{2}}$.  Its redshift ($z=0.063$) allows a
significant coverage of the cluster by the XMM-Newton field of view
($15^\prime$ in radius which corresponds to $1.48~{\rm
h_{50}^{-1}~Mpc}$ or about 0.5 times the cluster virial radius).  The
paper is organized as follows: after this introduction, Section 2
presents a methodology for spectro-imagery adapted to XMM background
and vignetting characteristics.  Our results in term of cluster global
properties and temperature profiles are presented in Section 3, with
particular emphasis on the various sources of errors.  Section 4
contains our conclusions.

\section{Data Analysis}
\label{sec:da}
A1795 was observed for $\sim 50$ ksec in Revolution 100 in Full Frame
Mode with the EPIC/MOS and pn camera.  We focus here on the MOS data. 
We generated calibrated event files with SASv4.1, except for the gain
correction.  Correct PI channels are obtained by interpolating gain
values obtained from the observations of the on-board calibration
source closest in time to A1795 observation.  Data were also checked
to remove any remaining bright pixels.  Spectra in various regions
were extracted to study temperature variations.  Only events
corresponding to detector regions in view of the sky are considered
(using SASv4.1).  The region corresponding to the bright point source
in the south was also excluded.  A proper treatment of the background
and vignetting effects are essential when analyzing extended sources
like cluster of galaxies.  This is discussed in the following two
sections.

\begin{figure}[t]
\epsfxsize=7.cm \epsfverbosetrue \epsfbox{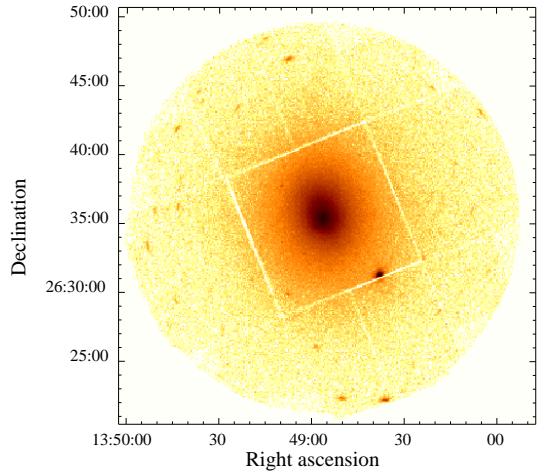} \caption[
]{{\footnotesize Combined EPIC/MOS1\&2 image of A 1795 in 
the $[0.3-10]\kev$ energy band  (logarithmic intensity).  }}
\label{fig:fima}
\vspace{-0.2cm}
 \end{figure}

\subsection{Background}
The XMM background shows considerable variations with time, with
flares of various durations and intensities.  As an example, we show
in Fig.~\ref{fig:frates} the count rate in the $[10-12]$ keV energy
band, where the emission is dominated by the background, due to the
small effective area of XMM-Newton/EPIC/MOS in this band.  It is
essential to remove these periods of high background, which are
induced by soft protons from solar flares collected by the telescopes. 
Not only is the S/N highly degraded, especially at high energy (where
the data are crucial for kT measurements), but it is impossible to
properly subtract the background, since its spectrum is varying with
time.  We thus removed all frames corresponding to a count rate
greater than 15 ct/100s in the $[10-12]$ keV band.  After this data
cleaning, the useful observing time for MOS1 and MOS2 is 32.1 ksec and
32.3 ksec, respectively.  The corresponding EPIC/MOS1\&2 image  in the $[0.3-10] \kev$ energy band is shown on
Fig.\ref{fig:fima}.  The remaining background is dominated by the
cosmic X--ray background at low energy and the Cosmic Ray (CR) induced
background at energies above typically 1.5 keV.

To estimate the background we used the Lockman Hole (LH) observations
made with the thin filter (Rev 70 and Rev 73 for MOS1 and Rev 71 for
MOS2).  Data were filtered for high background and regions
corresponding to the 10 brightest sources were
excluded\footnote{Background event files, combining several high galactic
latitude pointings, and thus allowing a statistically better estimate
of the CR induced background, were generated after the completion of
this work (D.Lumb, 2000) and was used in our analysis
of Coma (this issue).  The derived background is fully consistent
with the background used in this analysis}.  

\begin{figure}[t]
\epsfxsize=8.cm \epsfverbosetrue \epsfbox{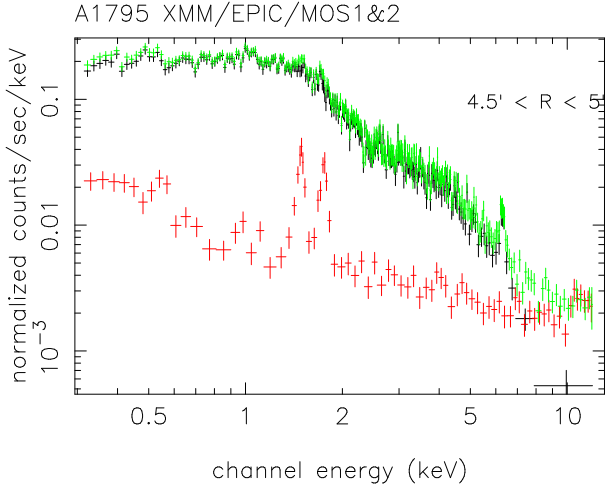} \caption[
]{{\footnotesize Summed EPIC/MOS1 and MOS2 spectra in the $4.5^\prime-5^\prime$
region.  Green points: raw spectrum.  Red points: the background
spectrum derived from the LH observation.  Black points: the
corresponding background subtracted cluster spectrum.} }
 \label{fig:fspecbkgr}
 \end{figure}

It is known that the CR induced background changes slightly in the FOV.
It is thus better to consider the same extraction regions in detector
coordinates for the source and the background.
Figure~\ref{fig:fspecbkgr} shows the raw cluster spectrum in the
$4.5^\prime-5^\prime$ region, the background spectrum derived from the
LH observation and the corresponding background subtracted
spectrum.  Note the Al and Si fluorescence lines in the
background spectrum and the very hard continuum above 2 keV typical of
CR induced background.  As a result, even for a bright cluster like
A1795 and relatively close to the center, the background dominates the
emission at high energies (here above 7 keV).  This is an important
limitation for temperature estimation.

The CR induced background changes slightly with time: the average
background level obtained after flare cleaning varies typically by less than
$10\% $ from observation to observation.  We considered two
criteria for background normalization (with each camera treated
independently): the count rate in the $[10-12]$ keV band in the FOV
(where sky X--ray emission is negligible) and the count rate in the
$[0.3-12]$ keV band in the region of the detector masked by the
proton shield (no sky emission).  The derived normalization factor
between the LH and A1795 observations varies from 0.94 to 1.03
depending on the camera and criteria considered.  In the following we
use a nominal a normalization factor of unity.  It will be varied by
$\pm 5\%$ to assess systematic uncertainties due to background
subtraction.

Note that this treatment allows a proper estimate of the CR induced
background, but not of the soft X-ray diffuse background (which varies
with position on the sky) and of the extra-galactic background (which
depends on the absorbing hydrogen column density along the line of
sight).  However, the source usually dominates the background except
at high energy (Fig~\ref{fig:fspecbkgr}) and we are mostly sensitive
to CR induced background.  Furthermore, $N_{H}$ values are similar in
the directions toward A1795 and LH and we checked from ROSAT survey
maps (Snowden \etal \cite{snowden1},\cite{snowden2}) that the X-ray
emission around the LH and A1795 in the soft band ($[0.4-1.2]$ keV)
differ by less than $30\%$ on average.

\subsection{Correction for vignetting effects}

\subsubsection{Methodology}

While the XMM-Newton/MOS spectral resolution does not show spatial
variations, the effective area at a given energy does depend on
position.  This effect has to be taken into account in modeling
spectra extracted from various regions.  The exact method is to
perform a convolution of a source model with the instrument response
(i.e. point by point) and to compare the resulting modeled spectrum
with the observed spectrum of the chosen region.  This method is
complex to implement, especially if the region under consideration is
not assumed to be isothermal.  In any case, it requires an a priori
model for the source spatial distribution.

A simpler and widely used method is to use the emission weighted
effective area in the region, derived from the {\it observed} global
photon spatial distribution (the same weighting is done at each
energy).  The incident source spectrum is simply compared with the
observed spectrum using this effective area (ARF).  This method has
severe drawbacks.  First, it introduces extra noise, since the
observed photon distribution used in the ARF average is noisy.  Taking
this error into account is not trivial.  Even more important, the
method introduces a {\it bias} in the determination of the spectral
parameters.  Since the effective area decreases with distance from the
center, the detected distribution of photons (at a given energy or
integrated over the energy band) is always more pronounced towards the
center than the source distribution (the well known distortion of the
observed image with respect to the incident image).  If the average of
the effective area is done using the observed distribution, too much
weight is given to the central regions.  The overall effective area is
thus overestimated.  Moreover since the variation with energy of the
effective area depends on position (the effective area decreases
more rapidly with distance for high energy photons than for low energy
ones), a bias is introduced.  The overestimate of the effective area
is higher at high energy than at low energy.  The resulting effect is
that the fit for the temperature is biased to lower values.

We propose instead the following method, which is easy to implement,
does not introduce bias and does not require any a priori assumptions
about the spatial variation of the source (either in intensity or in
spectroscopic properties).  When extracting the spectrum of a region
$Reg$, we weight each photon with energy $E_{I}$ 
falling at position $(x_{j},y_{j})$ 
by the ratio of the effective area at that position
$A_{x_{j},y_{j}}(E_{I})$ to the central effective area
$A_{0,0}(E_{I})$.  We can thus define a `corrected' spectrum C(I),
where C is the corrected photon counts in channel I centered on
$E_{I}$ with width $\Delta E_{I}$:
\begin{eqnarray}
C(I)& =&
 \sum_{j} w_{j};~(x_{j},y_{j}) \in Reg;  \\
 & &~~~~~~~~~~~~E_{I}-\Delta E_{I}/2<
E_{j} < E_{I}+\Delta E_{I}/2 \nonumber\\
w_{j}&=&\frac{A_{0,0}(E_{I})}{A_{x_{j},y_{j}}(E_{I})}
\label{eq:corspec}
\end{eqnarray}

This weighting factor has to be taken
into account in the error estimate. The variance on $C(I)$ is
\begin{equation}
\sigma^{2}(C(I)) = \sum_{j} w_{j}^{2}
\label{eq:corer}
\end{equation}

Note that this method corresponds to CORRECT mode in EXSAS for ROSAT
data reduction (Zimmermann \etal \cite{zimmermann}).  This `corrected'
spectrum is an estimate of the spectrum one would get if the detector
were flat.  It can thus be compared with the total spectrum of the
source given by the model, convolved with the instrument response at
the center of the detector.  The method is exact for an instrument
with perfect spatial and spectral ($\Delta E$) resolution, i.e., the
energy and position of the detected photons correspond exactly to
those of the emitted photons.  However, the method cannot introduce
significant bias as long as the vignetting effect is small at the
scale of the PSF and remains the same for energies differing by about
$\Delta E$.  This is certainly the case for XMM-Newton.  Its only
drawback, as compared to the direct exact method, is a degradation of
the statistical quality: the relative error is increased by a factor $
\sqrt{\langle{w_{j}^{2}}\rangle}/\langle{w_{j}}\rangle$ where the
brackets denote the average over the photons in a given energy bin I.
This is not a problem if the effective area does not vary much in the
region being considered
($\sqrt{\langle{w_{j}^{2}}\rangle}/\langle{w_{j}}\rangle\sim 1$).

For consistency,  the background
spectra were obtained using the same correction method as for the
source.  The CR-induced background component is not vignetted, but,
since we extract the background and source spectra from the same region
in detector coordinates, the correction factor is the same and thus
cancels.

\subsubsection{Vignetting data}
For the vignetting function of the telescope, we used the data
available on the XMM-Newton home page in August 2000, based on a
simplified model of the mirrors.  Data in the SASv4.1 CCF are wrong
and correct values derived from the last model implemented in SciSim
were only made available after the completion of this work.  We add
the vignetting effect due to the obscuration by the Reflection Grating
Array (RGA), as derived from SciSim (P.Gondoin private communication). 
This effect can be supposed to be energy independent but introduces an
azimuthal dependence in the vignetting function.  The overall vignetting is
in reasonable agreement with values derived from a comparison of on-
and off-axis (10' from centre) observations of the supernova remnant
G21.5$-$0.9 (Neumann, 2000).

\subsection{Spectral fitting}
The source spectra (with errors) computed as described above, are
binned so that the S/N ratio is greater than 3 $\sigma$ in each bin
after background subtraction.  The spectra are fitted with XSPEC using
a MEKAL model (Mewe \etal \cite{mewe1},\cite{mewe2}; Kaastra
\cite{kaastra}; Liedahl \etal \cite{liedahl}).  Since the spectra are
`corrected' we can use the on axis response file.  Note that the
response matrix we used (version v3.15) assumes no energy dependence
for the RGA transmission.  Only data above 0.3 keV are considered due
to remaining uncertainties in the MOS response below this energy.

\section{Results}

 \begin{figure}[t]
\epsfxsize=8.cm \epsfverbosetrue \epsfbox{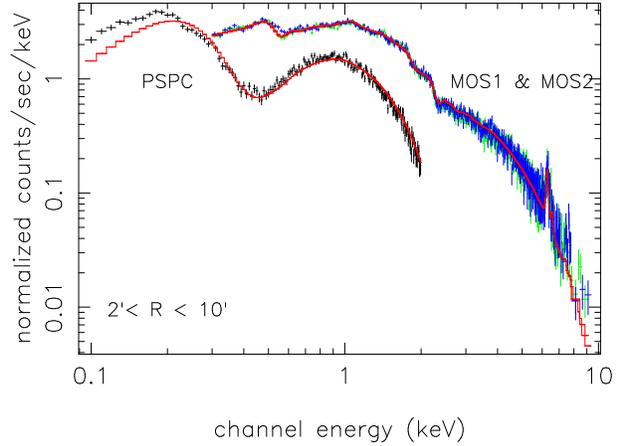}
\caption[ ]{{\footnotesize XMM and ROSAT spectra of the cluster
from the region covering radii of $2^\prime-10^\prime$.  Green (blue) points: 
EPIC/MOS1(2) data.  Black
points: PSPC data.  The EPIC spectra are background subtracted and
corrected for vignetting as described in Sec.~\ref{sec:da}.  The red
line is the best fit isothermal model  with ${\rm k}T= 5.95~\kev$,
 an abundance of 0.27 times the solar value and a relative normalization
 between the EPIC and PSPC data of 1.05.
}}
\label{fig:fpspcxmm}
\end{figure}

\subsection{Overall spectrum}

The overall MOS1 and MOS2 spectra, within $10^\prime$ of the cluster
center, excluding the central $R<2^\prime$ cooling flow region (see
below) is shown in Fig.~\ref{fig:fpspcxmm}.  The spectra are corrected
for vignetting and background subtracted as described above.  They are
compared with the PSPC spectrum of the same region.  As the PSPC FOV
is much larger, the background spectrum can be reliably estimated from
the outer region ($25^\prime-42^\prime$), free of cluster emission. 
The XMM-Newton and PSPC data are jointly fitted in the $[0.3-10]$ keV
range.  We let the relative normalization between the various
instruments be free but assumed a common temperature and abundance. 
When the $N_{\rm H}$ value is frozen to the 21 cm value ($N_{\rm H}=
1.05\times 10^{20}~{\rm cm^{-2}}$; Mittaz, Lieu \& Lockman \cite{mittaz}),
the best fit gives ${\rm k}T= 5.95\pm 0.1$~keV and an abundance of
$0.27 \pm 0.02$.  The reduced $\chi^{2}$ is $1.27$.  If the $N_{\rm
H}$ value is allowed to be free, we derive $N_{\rm H}=
0.95\pm0.07\times 10^{20}~{\rm cm^{-2}}$, in agreement with the 21 cm value
and the $\chi^{2}$ value, temperature and abundance are unchanged.  In
the following we thus fix the $N_{\rm H}$ value to the 21 cm value.

The reduced $\chi^2$ is reasonable, the deviations between model and
data are less than typically $10\%$\footnote{The most obvious misfit
is seen around the O edge (E$\sim0.55\kev$) in the MOS2 spectrum but
not in the MOS1 spectrum.  Its origin is unclear.  }.  One should also
note the excellent consistency between the MOS1 and MOS2 spectra
(relative normalization of 1.02) and between the PSPC and XMM
(relative normalization of 1.05) The best fit overall temperature in
this region is in good agreement with the global temperature value
from ASCA ($6.0\pm 0.3 \kev$, Markevitch \etal \cite{markevitch}) and
SAX ($6.0\pm 0.4 \kev$, Irwin \& Bregman \cite{irwin2}).  Overall this suggests
that the data processing and the on-axis response matrix are
reasonable.

When the best fit model is extrapolated down to 0.1 keV, the known
soft excess in the EUVE data below 0.2 keV (Mittaz, Lieu \& Lockman
\cite{mittaz}) is clearly apparent in the PSPC spectrum.  This excess
cannot be yet studied with XMM, due to the previously mentioned
calibration uncertainties (Section 2.3) in the response matrix in the
$[0.15-0.3]$ keV range.

\begin{figure}[t]
 \epsfxsize=8.cm \epsfverbosetrue \epsfbox{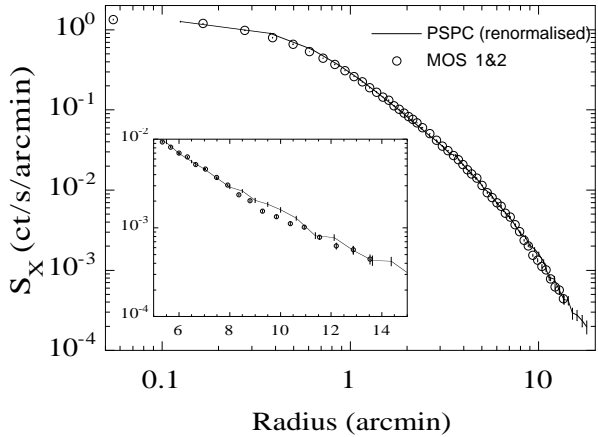} \caption[
 ]{{\footnotesize Comparison of XMM (open circles) and PSPC (full
 line) surface brightness profiles.  The profiles are background
 subtracted and corrected for vignetting (see text for details).  The
 PSPC data are renormalized to the total number of XMM counts in the $2^\prime$
 to $10^\prime$
 region.  Insert: zoom of the region where the discrepancy is maximum.
 Note the excellent agreement in shape. } }
\label{fig:fprof}
 \end{figure}

\subsection{Surface brightness profile}
 An independent check of the background subtraction and vignetting
 correction was performed by comparing XMM and PSPC surface brightness
 profiles.  The PSPC profile in the $[0.9-2.]$ keV energy range was
 extracted and corrected for vignetting using the standard EXSAS
 procedure for deriving exposure maps.  As was done in the spectral analysis, the
 background was estimated from outer regions in the FOV with point
 sources removed.  The PSPC profile is compared in Fig.~\ref{fig:fprof} with
 the vignetting-corrected summed MOS1 and MOS2 profile in
 the $[1-3]$ keV band.  The energy bands were chosen so that the relative
 emissivity between XMM and the PSPC were insensitive to possible
 temperature gradients, while at the same time retaining a good S/N ratio.  
 The XMM 
 background profile was estimated from the Lockman Hole data.  
 The profiles are binned so that there is at least a
 $5~\sigma$ detection in each bin.  For direct comparison
 the PSPC data in the figure were renormalized to the total XMM counts
 over radii from $2^\prime$ to $10^\prime$.  The agreement in shape is 
 excellent up to
 $15^\prime$ with the maximum difference (at the $25\%$ level only) 
 occurring at radii between $8^\prime$ and $11^\prime$.  This 
 comparison further validates our data  analysis method.
\begin{figure}[t]
\center
\epsfxsize=6.5cm \epsfverbosetrue \epsfbox{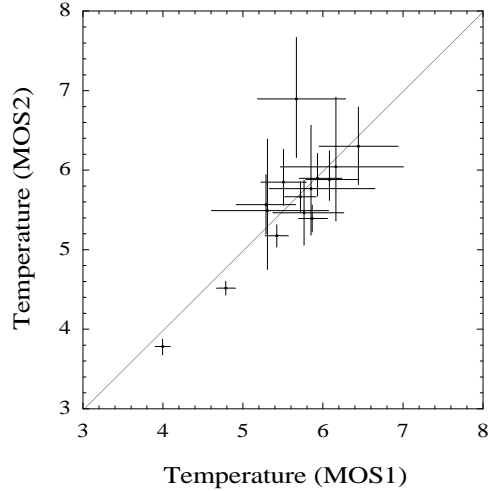} \caption[
]{{\footnotesize Comparison of the temperature derived from EPIC/MOS1
and MOS2 data for the various rings of the radial
temperature profile.  The errors are at the $90\%$ confidence level.
}}
\label{fig:fT2T1}
 \end{figure}

\subsection{Radial temperature profile}
Spectra in concentric rings (i.e., annuli) centered on the cluster X--ray emission
peak were extracted from MOS1 and MOS2 data.  We tried to attain
roughly similar precision in the temperature estimates from region to region.
Thus the widths of the various rings were chosen so
that at least a $5\sigma$ detection in the $[5-10]$ kev range was reached.
This was possible for all but the final annulus.  In addition a minimum 
width of $30^{\prime\prime}$ was set corresponding to the $90\%$ encircled 
energy radius of the PSF. Each vignetting-corrected and background-subtracted 
spectrum was fitted with an isothermal MEKAL model allowing the normalization, 
temperature and abundance to be the only free parameters.

\begin{figure}[t]
\epsfxsize=8.cm \epsfverbosetrue \epsfbox{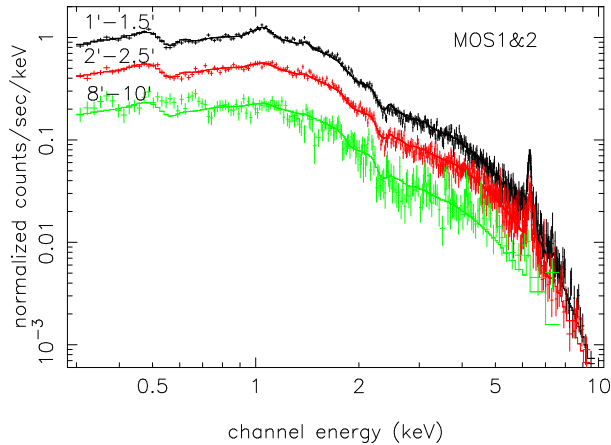} \caption[
]{{\footnotesize  Combined EPIC/MOS1\&2 spectra of
the inner $1^\prime-1.5^\prime$ region (black points), the $2^\prime-2.5^\prime$ region just outside the
cooling flow region (red points) and the outer $8^\prime-10^\prime$ ring (green points).
The full lines are the best fit isothermal model for each region. } }
\label{fig:f3spec}
\end{figure}

The temperatures determined from the MOS1 and MOS2 data are mutually consistent
(see Fig.~\ref{fig:fT2T1}).  We thus summed the MOS1 and MOS2
data and re-determined the temperature in each ring from the combined
data.  The fits are good with reduced $\chi^{2}$ values
in the range $0.98-1.24$. The worst $\chi^{2}$ value corresponds to the
central rings, where we are affected by the cooling flow.
Typical spectra are shown in Fig.~\ref{fig:f3spec}:
the inner $1^\prime-1.5^\prime$ region, the $2^\prime-2.5^\prime$ region just outside the
cooling flow region and the last ring ($8^\prime-10^\prime$ region).  The 
errors in our temperature estimates consist of the quadratic sum of
the statistical error (at 
the $90\%$ confidence level)  and systematic errors estimated by 
varying the background level by $\pm5\%$. 
We also fitted the spectra in the restricted $0.3-7$
keV range, which is less sensitive to background and we did not obtain any
significant differences in the derived temperatures.
We also tentatively estimate the temperature in an outer region covering
radii of $10^\prime-12^\prime$.
The temperature from this region is uncertain and the value should be used with caution. The fit is not
good ($\chi^{2}=1.8$), due to an excess at energies above 5 kev and
a poor fit at low energies. Both effects are likely due to uncertainties in the
background (both the X-ray background at low energies and CR-induced
background at high energies) which dominates the emission. The error
includes systematic errors estimated by varying the background level
by $\pm 5\%$ and restricting the fit to the $[0.3-5]$ keV energy range.

\begin{figure}[t]
\epsfxsize=8.cm \epsfverbosetrue \epsfbox{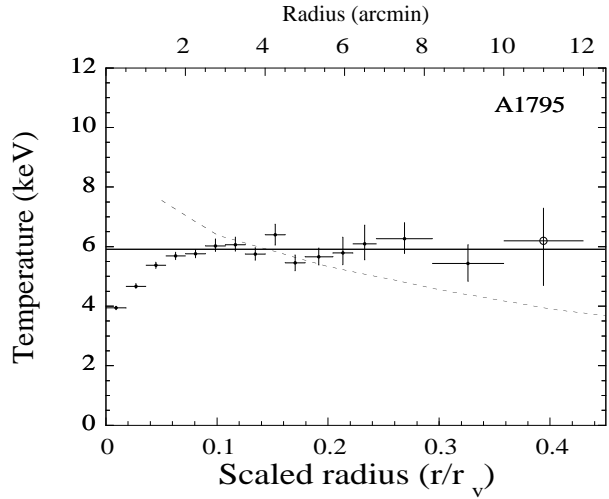}
\caption{{\footnotesize Radial temperature profile as a function of
angular radius (upper axis) and scaled radius (lower axis), derived
from XMM/EPIC data.  The horizontal line corresponds to the mean value
derived from fitting the overall spectrum of the region from
 $2^\prime$ to $10^\prime$ radius.  The dotted line is the universal
profile derived by Markevitch \etal (\cite{markevitch}) from ASCA data,
normalized to the virial radius and mean temperature of A1795.  }}
\label{fig:fkTprof}
\end{figure}

The resulting temperature profile is shown in Fig.~\ref{fig:fkTprof}
as a function of radius both in physical units (upper axis) and 
scaled by the virial radius (bottom axis).  The scaled radius is defined as
r/$r_{200}$ where $r_{\rm 200}=2.76\,{\rm Mpc}$ is the virial radius for a
$6\kev$ cluster  at $z=0.063$ (Evrard \etal \cite{evrard}).

\section{Conclusion}

We have determined the temperature profile of A1795 up to a typical
radius 0.4 times the virial radius.  Both vignetting effects and
background subtraction are reasonably well understood.  The
temperature profile can be compared with the profile derived by SAX
(Irwin \& Bregman \cite{irwin2}) and ASCA (Markevitch \etal
\cite{markevitch}).  The dramatic improvement in accuracy and spatial
resolution of the profiles must be noticed.  However, our results show
that the limiting factor for determining temperature profiles with XMM
is the high level of the CR-induced background.  Improvement of
background models would be of great help.

The drop of temperature in the center as measured by ROSAT (Briel and
Henry 1996), but not seen with SAX, is unambiguously confirmed.  The
drop in temperature is also readily apparent in the spectrum of the
central region shown in Fig.\ref{fig:f3spec}, which shows the
increasing importance of the Fe L line complex.  Moving in toward the
cluster center the temperature starts to drop at about $2^\prime$;
further in the temperature profile is, for the first time, well
resolved.  A significant excess in the surface brightness profile (as
compared to the overall $\beta$ model) is also observed in the same
region (Briel \& Henry \cite{briel2}).  This temperature drop
indicates that the gas is cooling.  A detailed study of this inner
region, based on both EPIC and RGS data, can be found in Tamura \etal
(\cite{tamura}).

Beyond $2^\prime$ the temperature results are consistent with an
isothermal profile at the mean temperature (full line in
Fig.\ref{fig:fkTprof}) in agreement with SAX results.  The ASCA
temperature beyond $6^\prime$ (Markevitch \etal \cite{markevitch}) is
also in excellent agreement with our data, but is significantly higher
in the $1.5-6'$ radial range (k$T =8.2^{+1.6}_{-1.5}\kev$).  The
temperature in the central $r<1.5'$ region is corrected for the
presence of the cooling flow and cannot be directly compared with our
data.  We also show in Fig.\ref{fig:fkTprof} the decreasing
`universal' profile obtained by Markevitch \etal (\cite{markevitch})
from their compilation of ASCA temperature profiles (dotted line).  It
can be safely compared to our data outside the cooling flow region. 
Our results do not support this type of profile for A1795.  However,
we emphasize that no definitive conclusion should be drawn yet, based
on the observation of one cluster only and while we are still in the
early phases of the XMM mission.  In particular we have assumed that
the (small) energy dependence in the shape of the PSF can be
neglected.  This still needs to be confirmed with in-flight
calibration data and detailed simulations.

\begin{acknowledgements}
We would like to thank J. Ballet for support concerning the SAS
software and useful discussions on the vignetting correction method.
We thank J.-L. Sauvageot for providing the gain correction.
\end{acknowledgements}


\begin{thebibliography}{}


\bibitem[1992]{briel1} Briel U. G., Henry J. P., B\"ohringer H., 1992, A\&A,  259, L31

\bibitem[1996]{briel2} Briel U.G., Henry J.P., 1996, ApJ, 472, 131

\bibitem[1999]{horner} Horner D.J., Mushotzky R.F., Scharf C.A., 1999, ApJ, 520, 78

\bibitem[2000]{ettori}Ettori S., Bardelli S., De Grandi S., Molendi S., 
Zamorani G., Zucca E., 2000, MNRAS, 318, 239

\bibitem[1996]{evrard} Evrard A.E., Metzler C.A., Navarro J.F., 1996, ApJ, 469, 494

\bibitem[1999]{irwin1} Irwin J.A., Bregman J.N., Evrard A.E., 1999, ApJ, 519, 518

\bibitem[2000]{irwin2} Irwin J.A., Bregman J.N., 2000, ApJ, 538, 543

\bibitem[2001]{jansen} Jansen, F., Lumb, D., Altieri, B. \etal 2001,
A\&A, 365 (this issue)

\bibitem[1992]{kaastra} Kaastra J.S., 1992, An X-ray spectral Code for Optically
Thin Plasmas, Internal SRON-Leiden Report, updated version 2.0

\bibitem[2001]{tamura} Tamura T., Kaastra J.S., Petreson J.R.,  \etal 2001,
A\&A, 365 (this issue)

\bibitem[1995]{liedahl}  Liedahl, D.A., Osterheld, A.L., and Goldstein, W.H. 1995, ApJL,
438, 115

\bibitem[1998]{markevitch} Markevitch M., Forman W.R., Sarazin C.L., Vikhlinin A.,
 1998, ApJ, 503, 77

\bibitem[1985]{mewe1}  Mewe, R., Gronenschild, E.H.B.M., and van den Oord, G.H.J. 1985,
       A\&AS, 62, 197

\bibitem[1986]{mewe2} Mewe, R., Lemen, J.R., and van den Oord, G.H.J. 1986, A\&AS, 65,
511

\bibitem[2000]{molendi} Molendi S., De Grandi S., Fusco-Femiano R., ApJ, 2000, 
533, L43

\bibitem[1998]{mittaz} Mittaz J., Lieu R., Lockman F., 1998, ApJ 498, L17

\bibitem[1997]{NFW} Navarro J.F., Frenk C.S., White S.D.M., 1997, ApJ, 490, 493

\bibitem[1999]{neumann1} Neumann D.M., Arnaud M., 1999, A\&A, 348, 711

\bibitem[2000]{neumann2} Neumann D.M., 2000, Report on the in-flight vignetting
calibration of the MOS cameras aboard the XMM NEWTON satellite, CEA/Saclay,
France {\tt http://xmm.vilspa.esa.es/calibration/}

\bibitem[1997]{oukbir} Oukbir J., Blanchard A., 1997, A\&A, 317, 1

\bibitem[1999]{ponman} Ponman T.J., Cannon D.B., Navarro J.F., 1999, Nature, 397, 135;

\bibitem[1995]{snowden1} Snowden S.L. , Freyberg  M. J., Plucinsky  P. 
P. \etal  1995, ApJ, 454, 643

\bibitem[1997]{snowden2} Snowden S.L., Egger R., Freyberg M. J. \etal,
1997, ApJ, 485, 125

\bibitem[2001]{turner} Turner, M.J.L., Abbey, A., Arnaud, M. et al.
2001, 
A\&A, 365 (this issue)


\bibitem[1993]{white1} White S.D.M., Navarro J.F., Evrard A.E., Frenk C.S., 1993,
Nat. 366, 429

\bibitem[2000]{white2} White D.A., 2000, MNRAS, 312, 663

\bibitem[1998]{zimmermann} Zimmermann U., Boese G., Becker W., Belloni T., D\"obereiner S.,
Izzo C., Kahabka P., Schwentker O., 1998, EXSAS User's Guide
MPE Report  ROSAT Scientific Data
Center

\end{thebibliography}
\end{document}